\documentstyle[12pt]{article}
\newcommand{\beq}{\begin{equation}}
\newcommand{\eeq}{\end{equation}}

\def\quaft{{\textstyle {{{1}\over{4\pi\alpha'}}} }}
\def\half{{\textstyle{1\over2}}}

\def\p1half{{\textstyle{{{p+1}\over{2}}}}}

\def\23phalf{{\textstyle{{{23-p}\over{2}}}}}

\begin{document}
\thispagestyle{empty}
\begin{titlepage}

\bigskip
\hskip 3.7in{\vbox{\baselineskip12pt
}}

\bigskip\bigskip
\centerline{\large\bf 11d Electric-Magnetic Duality and the Dbrane Spectrum}

\bigskip\bigskip
\bigskip\bigskip
\centerline{\bf Shyamoli Chaudhuri \footnote{Email:
shyamolic@yahoo.com} } \centerline{1312 Oak Drive}
\centerline{Blacksburg, VA 24060}
\date{\today}

\bigskip
\begin{abstract}
\vskip 0.1in\noindent
We consider the {\em gedanken} calculation of
the pair correlation function of 
spatially-separated macroscopic string solitons in 
strongly coupled type IIA string/M theory, with the
 macroscopic strings wrapping the eleventh
dimension. The 
supergravity limit of this correlation function
with well-separated, {\em pointlike}
macroscopic strings 
corresponds to having also taken the IIA string coupling constant to 
zero. Thus, the pointlike limit of the {\em gedanken} correlation
function can be given a precise worldsheet description 
in the 10D weakly-coupled 
type IIA string theory, analysed by us in hep-th/0007056
[Nucl.\ Phys.\ {\bf B591} (2000) 243].
The requisite type IIA string amplitude is the supersymmetric 
extension of the worldsheet formulation of an 
off-shell closed string tree propagator in bosonic string theory,
a 1986 analysis due to Cohen, Moore, Nelson, and Polchinski.
We point out that the evidence for pointlike sources of the 
zero-form
field strength provided by our
worldsheet results clarifies that the 
electric--magnetic duality in the Dirichlet-brane spectrum 
of type II string theories is {\em eleven}-dimensional. 
\end{abstract}

\end{titlepage}

\section{Introduction}

\vskip 0.1in The discovery of Dirichlet pbranes in the 10d type II
string theories \cite{dbrane}, with $-1$ $\le$ $p$ $\le$ $9$,  
also raises a puzzle for 10d electric-magnetic
duality; let us begin by explaining this 
puzzle. Recall that Dbranes are domain wall topological defects
in an embedding target spacetime that are found to occur naturally
among the ground states of type I, and type II, supersymmetric
open and closed string theories \cite{dbrane}. In addition to
being charged under an antisymmetric (p+2)-form field strength,
Dpbranes carry worldvolume Yang-Mills gauge fields. An important
distinction from generic solitonic solutions of classical gauge,
or supergravity, theories is that string perturbation theory at
weak coupling is well-defined in the background of the infinitely
thin Dpbrane soliton \cite{witncom}. Thus, the quantum scattering
amplitudes of Yang-Mills gauge theory in the background of a
pbrane soliton of {\em zero width} can be obtained from the low
energy gauge theory limit of corresponding perturbative string
amplitudes computed in the background of a Dpbrane
\cite{smallinst}. This long-unsolved problem in gauged soliton
dynamics cannot be addressed by standard field theoretic
techniques.

\vskip 0.1in The key property of supersymmetric open and closed
string theories that enables the solution of a problem which is
surprisingly clumsy to formulate within the semi-classical
collective coordinate formalism is the existence of target
spacetime duality transformations linking pairs of string vacuum
states: $R$ $\to$ $R^{\prime}$ $=$ $\alpha^{\prime}/R$,
interchanging target spactimes with large, and small, radius.
Thus, the existence of vacua with Dirichlet-brane target spacetime
geometries can be straightforwardly established by performing a
sequence of T-duality transformations on the standard vacuum with
the 10D flat spacetime Lorentzian geometry
\cite{dbrane}. Moreover, this description is {\em
exact}: it is not necessary to resort to the complicated
alternative that describes a Dbrane soliton as composed out of the
collective modes of an ensemble of, more fundamental, open and
closed strings.

\vskip 0.1in An important step forward in our understanding of the
significance of Dbranes is provided by Polchinski's identification
of Dbranes as the BPS states of the type II supergravities
carrying Ramond-Ramond (R-R) charge \cite{dbrane}. This leap of
insight establishes contact between the worldsheet formulation of
infinitely thin domain wall pbrane solitons carrying Yang-Mills
gauge fields and the zero width limits of a known family of type
II classical supergravity soliton solutions \cite{pbrane}. R-R
solitons contain worldvolume gauge fields and, for overlapping
Dbranes, the gauge group is nonabelian. This implies a
noncommutative structure for spacetime at short distances
\cite{witncom}, an insight that follows straightforwardly from the
worldsheet formulation \cite{dbrane}. Most importantly, as becomes
clear with the collective insights assembled together in the
landmark work \cite{witdual}, the R-R sector solitons are the
precise BPS states known to be of significance in establishing the
conjectured strong-weak coupling duality relations linking the
different backgrounds of String/M theory.

\vskip 0.1in By application of a sequence of T-duality
transformations on the standard ten-dimensional flat spacetime
Lorentzian vacuum we can infer, therefore, a spectrum of Dpbranes
with $p$ in the range: $-1$ $\le$ $p$ $\le$ $9$, covering
D-instantons thru D9branes \cite{dbrane}. This raises the
following puzzle. By a generalization of the Dirac quantization
condition for electric and magnetic point charges in four
spacetime dimensions, one would expect that the product of the
quantum of charge for a Dpbrane, and that of its $d$-dimensional
Poincare dual D(d-p-4)brane, satisfies the relation:
\begin{equation}
\nu_p \nu_{d-4-p} = 2\pi n , \quad \quad n \in {\rm Z}  \quad .
\label{eq:diracq}
\end{equation}
The spectrum of values for $p$ listed above does not cover all of
the expected charges when d=10: we are apparently
missing a \lq\lq D(-2)brane" and a
\lq\lq D(-3)brane", the 10d Poincare duals of the D8brane and D9brane. This 
is especially puzzling because, in his
groundbreaking work \cite{dbrane}, Polchinski has shown that the
quantum of Dpbrane charge could be computed from first principles
using the worldsheet formalism, predicting also that the value of
$n$ in the Dirac quantization relation is {\em unity}. Thus,
although we were expecting to find evidence for electric-magnetic
duality in the full Dpbrane spectrum we appear, instead, to have
found a direct clash with Poincare-Hodge duality in ten target spacetime
dimensions.

\vskip 0.1in We will show in this paper that there is a simple
resolution to this puzzle, and the evidence in favor of it can be
convincingly found in an extension of Polchinski's worldsheet
calculation of the Dpbrane tension \cite{dbrane}. In particular,
we will argue that the Dirichlet-pbrane spectrum of the type II string
theories actually contains {\em twelve} elements:
$-1$ $\le$ $p$
$\le$ $9$, plus the pointlike sources of zero-form field strength,
in full agreement with Poincare-Hodge duality in eleven
target spacetime dimensions. The additional Dirichlet-brane is
the source for the scalar field strength, $F_0$, appearing in
Roman's massive 10d type IIA supergravity theory \cite{romans};
$F_0$ is the Poincare dual of the 10-form field strength coupling
to Polchinski's D8brane \cite{dbrane}. In section 2
of this paper, we will begin by reviewing some of the subtleties in
finding evidence for solitonic target spacetime 
sources for the zero-form field strength, 
 using standard 10d
supergravity, and superstring, techniques. In section 3, we 
formulate our {\em gedanken}
calculation in 11d strongly coupled IIA string/M theory, elucidating the
nature of the worldsheet evidence for pointlike sources of the Romans
zero-form field strength given by us in \cite{flux}. The details of the worldsheet 
analysis appear in
the appendices of the paper. Our conclusions in section 4 
focus on the implications
of our results for nonperturbative string/M theory.

\section{The Magnetic Dual of D8brane Charge}

\vskip 0.1in It is helpful to begin by considering the
ten-dimensional type IIA, and type IIB, supersymmetry algebras. We
remind the reader that Poincare duality in $d$ spacetime
dimensions \cite{teit,dbrane,polbook} relates a (p+2)--form field
strength, $F_{p+2}$, to its dual field strength, $F_{d-2-p}$ $=$
$*F_{p+2}$, of rank $d-p-2$. A priori, the rank can take any value
in the range $0$ to $d$. Supersymmetry places additional
restrictions and so, in ten dimensions, for example, we have $p+2$
$=$ $0,2,4,6,8,10$ in the non-chiral type IIA algebra, and $p+2$
$=$ $1,3,5,7,9$ in the chiral type IIB algebra. Each value of $p$
determines a different central extension of the 10d $N=2$
supersymmetry algebra with 32 supercharges. Thus, in the IIA
algebra, we have:
\begin{eqnarray}
\{ Q_{\alpha} , {\bar{Q}}_{\beta} \} =&&  \left [ P_{\mu} + (2\pi
\alpha^{\prime })^{-1} Q_{\mu}^{\rm NS} \right ]
\Gamma^{\mu}_{\alpha \beta}  \cr \{ {\tilde{Q}}_{\alpha} ,
{\tilde{{\bar{Q}}}}_{\beta} \} =&& \left [ P_{\mu} - (2\pi
\alpha^{\prime })^{-1} Q_{\mu}^{\rm NS} \right ]
\Gamma^{\mu}_{\alpha \beta} \cr \{ Q_{\alpha} ,
{\tilde{{\bar{Q}}}}_{\beta} \} =&&  \sum_{p=0}^{8}
{{\tau_{p}}\over{p!}} Q^{\rm R}_{\mu_1 \cdots \mu_{p} } \left (
\Gamma \Gamma^{\mu_1} \cdot \Gamma \Gamma^{\mu_2} \cdots \Gamma
\Gamma^{\mu_{p}} \right )_{\alpha \beta} \quad . \label{eq:susy}
\end{eqnarray}
Here, $p$ takes only even values, and $\mu$ runs over all ten
target spacetime coordinate labels, the $\{ \Gamma^{\mu}, \Gamma =
\Gamma^0 \Gamma^1 \cdot \Gamma^{9} \} $ are the Dirac matrices in
the 10d Clifford algebra, and the $Q_{\alpha}$,
${\tilde{Q}}_{\alpha}$, $\alpha$$=$$1$, $\cdots$, $16$ are,
respectively, left-moving and right-moving supercharges. The
anticommutator of any two chiral supercharges closes on the
generator of 10d spacetime translations, and on a central
extension in the Neveu-Schwarz (NS) sector: the fundamental
string, which couples to the NS sector two-form potential.
Nominally, one might expect that the sum on the right-hand-side of
the last equation could begin with $p$ $=$ $0$: it is known that
the massive type IIA supergravity action contains both a ten-form
field strength \cite{romans}, 
as well as its Poincare dual scalar field strength \cite{bergt}.
Could we be missing an additional contribution to the right-hand-side
of this equation representing the charge of a Dirichlet (-2)brane?

\vskip 0.1in It turns out that there is no conflict with the ten-dimensional 
algebra of supercharges given above because the missing Dpbrane of interest 
will turn out to be of {\em eleven} dimensional origin: the algebra above contains only 
ten translation generators, 
whereas the strongly coupled type IIA string/M theory is 
eleven dimensional. Notice, also, that the supersymmetry algebra of relevance 
is that describing the strong coupling limit of Romans' {\em massive} type IIA supergravity. 
This is not the same thing as 11d supergravity, the strong coupling limit of the
{\em massless} type IIA supergravity \cite{west,hulltown,witdual}, and so we must
invoke the supersymmetry algebra of nonperturbative string/M theory itself. Namely,
for clear-cut algebraic evidence of sources for the
 zero-form field strength, we require a 
framework incorporating
both the weak, and strong, coupling limits of the massive type IIA string theory. We will 
return to this point in the conclusions.

\vskip 0.1in Next, let us move on to consider some aspects of the 
correspondence between the 10d type IIA
superstring theory, and its low energy 10d type IIA supergravity limit. Recall 
that, for any value of $p$, the flux quantum, $\nu_p$, differs from
the quantum of Dpbrane charge, $\mu_p$, and the physical
Dpbrane-tension, $\tau_p$, in its lack of dependence on the closed
string coupling, $g$, and fundamental string tension,
$\alpha^{\prime -1/2}$. As a consequence, it can be calculated
unambiguously in {\em weakly-coupled} perturbative string theory
\cite{dbrane}. Moreover, Polchinski showed that the flux quanta of 
Dirichlet pbranes could be computed from
first principles by worldsheet methods \cite{dbrane}, satisfying the Dirac
quantization condition given above with $n$ equal to unity. Namely, 
we have the
relations:
\begin{equation}
\kappa^2 \tau_p^2 \equiv g^2 \kappa_{10}^2 \tau_p^2 =
\kappa_{10}^2 \mu_p^2 , \quad\quad \kappa_{10}^2 \equiv \half
(2\pi)^7 \alpha^{\prime 4} \quad . \label{eq:charge}
\end{equation}
Here, $\kappa$ is the physical value of the ten-dimensional
gravitational coupling. and the dimensionless closed string
coupling, $g$ $=$ $e^{\Phi_0}$, where $\Phi_0$ is the vacuum
expectation value of the dilaton field. The physical Dbrane
tension, $\tau_p$, is the coefficient of the worldvolume action
for the Dpbrane \cite{dbrane}. Finally, recall that the flux
emanating from a Dirichlet pbrane source is obtained by
integrating $F_{p+2}$ over a sphere in $(p+2)$ dimensions that
encloses the source:
\begin{equation}
\int_{S_{p+2}} F_{p+2} = 2 \kappa_{10}^2 \mu_{p} = \nu_p \quad .
\label{eq:flux}
\end{equation}
The spacetime Lagrangian for the bosonic sector of Romans' 10D 
massive IIA supergravity theory was obtained in \cite{bergt} . Note
that both the ten-form
field strength, as well as its Poincare dual scalar field strength
appear explicitly in the spacetime Lagrangian
\cite{bergt,polbook}. In the Einstein frame metric, the
bosonic part of the massive IIA action takes the simple form
\cite{bergt,polbook}:
\begin{equation}
S = {{1}\over{2\kappa^2}} \int d^{10} X {\sqrt{-G}}
  \left ( R_G - \half (\partial \Phi)^2  - \half |H_3|^2 - \half e^{5\Phi/2} M^2 \right )
+  {{1}\over{2\kappa^2}} \int d^{10} X {\sqrt{-G}} M F_{10} \quad
. \label{eq:acty}
\end{equation}
where $M$ is an auxiliary field to be eliminated by its equation
of motion. $F_{10}$ is the non-dynamical ten-form field strength,
which can be dualized to a zero-form, or scalar, field strength,
$*F_{10}$. This non-dynamical constant field generates a uniform
vacuum energy density that permeates the ten-dimensional
spacetime, thus behaving like a cosmological constant: $F_{10}$
$=$ $d C_9$, varying with respect to the R-R gauge potential,
$C_9$, gives $M$ $=$ constant, and varying with respect to $M$
gives, $F_{10}$ $=$ $ M e^{5\Phi/2} V_{10}$, where $V_{10}$ is the
volume of spacetime. Thus, we can identify the dualized scalar
field strength with the mass parameter: $*F_{10}$ $=$ $M
e^{5\Phi/4}$. Polchinski identified the D8brane as the source for
this ten-form field strength in \cite{dbrane}, and a solution to the beta function
equations with the identifiable spacetime geometry and background
fields of the D8brane was subsequently discovered
\cite{bergt,pbrane}. Can we likewise identify a classical soliton 
of the 10D massive IIA supergravity theory that can be interpreted
as a point source of the zero-form field strength? 

\vskip 0.1in Since the worldvolume of a
D(-2)brane is \lq\lq (-1)-dimensional",  it is obvious that 
we cannot look for a 
suitable solitonic solution to the beta function equations of the 
massive type IIA theory Ramond-Ramond
sector with identifiable worldvolume spacetime geometry,  
as was done for the remaining members of the Dbrane spectrum
\cite{dbrane,pbrane}. Instead, we will infer the existence of 
sources for the scalar field strength by calculating the long-range
force between two pointlike defects in a solitonic target space
geometry. These pointlike defects should not be confused with 
Dinstantons, a point that will become clear in the next section.
Notice that our calculation {\em is} a precise analog of 
Polchinski's calculation of 
the long-range
force between generic Dpbranes, because the Dpbranes are always 
\lq\lq pointlike" from the perspective of the observor in the bulk 
target spacetime, for any value of $p$ 
\cite{dbrane}. It is only the precise power law behavior of the
long-range force that identifies the extended worldvolume of the 
Dpbranes, $R^{-(7-p)}$ for a pair of Dpbranes. Thus, for point
sources of the zeroform field strength, we are looking for a $R^{-9}$ fall-off
in the long-range force. 

\vskip 0.1in 
It turns out that  
there exists a fundamental 
string soliton in the massive 10D type IIA 
supergravity with the requisite target spacetime 
geometry that provides a helpful analog for our {\em gedanken}
calculation in strongly coupled type IIA string/M theory in the 
next section \cite{flux}. Soliton solutions to the beta function equations of
the 10D massive IIA supergravity in the D0-D8-F1 sector have 
been exhaustively studied in \cite{troost,jmo,bergc}. Consider a fundamental
string terminating at a point within the worldvolume of a D8brane
\cite{ps}. Generically, this configuration will break $1/4$ of the
supersymmetries of the type IIA string. Since the Killing spinor
is annihilated by the projections of both D8brane and fundamental
string, we would infer that the intersection of the worldvolumes
of the D8brane and F1 string is a D0brane. This is because the
product of either two projection operators gives the third:
\begin{equation}
{\hat{\Pi}}_{D8} = \half (1+ \Gamma^{9}) , \quad
{\hat{\Pi}}_{F1} = \half (1+ \Gamma^{09}) , \quad
{\hat{\Pi}}_{D0} = \half (1+ \Gamma^{0}\Gamma_{11} )  \quad .
\label{eq:projs}
\end{equation}
The observation that the termination of a fundamental string in
the worldvolume of a Dpbrane, with all $p$$+$$1$ translation
invariances broken, should act like a point source of the magnetic
charge dual to the Dpbrane charge, is originally due to Polchinski
and Strominger \cite{ps}. There exists, not surprisingly, a
classical soliton solution of the massive IIA supergravity with
precisely these properties among the soliton solutions of the
D0-D8-F1 sector of the massive IIA supergravity theory analyzed by
Massar and Troost \cite{troost}, \cite{troost}, and independently
by Janssen, Meessen, and Ortin \cite{jmo}. We will describe
the target spacetime 
geometry as given in the paper by Massar and Troost
\cite{troost,flux}. These authors found a
new solution to the string beta function equations at leading
order in the $\alpha^{\prime}$ expansion which describes the
following spacetime geometry: a fundamental string extending in
the direction, $X^9$, orthogonal to the worldvolume of an D8brane
carrying, in addition, $C_1$, and $B_2$, fields. With the notation
$z$$=$$X^9$, $t$$=$$X^0$, the background fields take the form:
\begin{eqnarray}
ds^2 =&& -H^{1/8} h^{-13/8} dt^2 + H^{9/8} h^{-5/8} dz^2  +
H^{1/8} h^{3/8} ((dr)^2 + r^2 (d\Omega_7)^2 )
\nonumber\\
  H(z) =&& c + M z , \quad h(r) = 1 + {{Q}\over{r^6}}
\nonumber\\
e^{\Phi} =&& H^{-5/4}h^{1/4} , \quad B_{tz} = - h^{-1} ,
\quad C_z = H h^{-1} \quad .
\label{eq:mettrs}
\end{eqnarray}
The fundamental string extends in the direction perpendicular to
the worldvolume of the D8brane, generically breaking $1/4$ of the
supersymmetries. A nice discussion of this soliton and its
extension under $SL(2,Z)$ to a whole multiplet of $(p,q)$ soliton
strings, first pointed out by us in \cite{flux}, has appeared
recently in \cite{bergc}. 

\vskip 0.1in We will show in the next section that,
from the perspective of an observer in the worldvolume of the
D8brane, the endpoint of the fundamental string soliton
can behave like a
point source of Dirichlet (-2)brane charge. The 10D
observor can therefore measure the zero 
string coupling remnant of a calculation that, strictly speaking,
 belonged in the 11D strongly
coupled type IIA string theory. 
The massive fundamental string soliton of interest to us 
will wrap the orthogonal spatial coordinate
$X^{10}$,
rather than $X^9$, and its pointlike limit, $R_{10}$ $\to$ $0$, 
yields physics that is accessible in the low energy
limit of the 10D weakly coupled type IIA string theory. This
is the key point that enables a precise calculation of the 
{\em tension} of the pointlike source of zeroform field strength using standard
type IIA worldsheet techniques \cite{ferm,flux}.

\section{Pointlike Sources of Magnetic D(-2)brane Charge}

\vskip 0.1in The motivation for our {\em gedanken} calculation
came from an analogous worldsheet calculation in type II string
theory. If pointlike sources of the zeroform field strength 
exist, it would clearly be
helpful to find a first-principles calculation of the {\em
tension} of such pointlike sources in the worldsheet formalism,
analogous to Polchinski's well-known analysis for the remaining
Dpbranes, $-1$ $\le$ $p$ $\le$ $9$ \cite{dbrane}. It turns out that
the answer is to be found in the relatively poorly-explored
worldsheet formalism for {\em macroscopic} loop amplitudes in the
perturbative string theories \cite{cmnp,wils,ferm}, as was pointed
out by us in \cite{flux}.

\vskip 0.1in The worldsheet formulation of macroscopic loop
amplitudes, starting from an extension of the standard Polyakov
string path integral, was first explored by Cohen, Moore, Nelson,
and Polchinski \cite{cmnp}. These authors derived an expression
for a covariant off-shell closed string tree propagator in bosonic
string theory: the expectation value for a closed bosonic string
to propagate from a fixed loop, ${\cal C}_i$, in the embedding
target spacetime to a different fixed loop, ${\cal C}_f$, thru a
spatial distance $R$. Ref.\ \cite{cmnp} limited their discussion
to the case of pointlike boundary loops. In recent works by
Chaudhuri, Chen, and Novak, the analysis in \cite{cmnp} has been
extended to the case of macroscopic boundary loops in both the
bosonic \cite{wils}, and type II supersymmetric \cite{ferm,flux},
string theories, and including the results for target spacetime
backgrounds with an external two-form field strength. In this
section, we will begin by reviewing this formalism, explaining why
the factorization limit of the macroscopic loop amplitude in type
I$^{\prime}$ theory yields an expression for the
tension. As a byproduct of our analysis, we will also
succeed in computing the long range force between a pair of
pointlike sources of zeroform field strength in the presence of an external two-form field
strength, a result that appears in \cite{flux,ferm}.

\vskip 0.1in It is helpful to clarify precisely where our analysis
will differ from Polchinski's computation of the force between two
pointlike spacetime events, namely, a pair of Dinstantons. Recall
that the Dpbrane tension, $-1$ $\le$ $p$ $\le$ $9$, was extracted
from a computation of the graviton-dilaton one-point function on
the disk lying in the worldvolume of the Dpbrane. A simple trick,
exploited in \cite{dbrane}, that gives an unambiguously normalized
disk one-point function, is to extract the one-point function from
the factorization limit of the annulus graph, with boundaries
lying in the worldvolume of a pair of parallel, static Dpbranes.
For the case of pointlike Dinstantons, the annulus amplitude is
computed with all ten embedding target spacetime coordinates
obeying the Dirichlet boundary condition \cite{dbrane}.

\vskip 0.1in
Now consider the anomaly-free and perturbatively renormalizable
weakly coupled 10d type IIA string theory in the
background with 32 D8branes. One of the two supersymmetries of
the IIA string has been broken by the presence of orientifold
planes at $X^9$$=$$0$, and $X^9$$=$$R_9$, and all 32 D8branes
lie in the orientifold plane at the origin. A
T-duality transformation maps this background to
an analogous background with 32 D9branes in the 10d type IIB
string theory. The stack of coincident Dpbranes carries 
worldvolume Yang-Mills gauge fields 
with gauge group $SO(32)$. At finite string coupling, an eleventh
target space \lq\lq dimension" emerges, corresponding to the vacuum expectation
value of the scalar dilaton field \cite{witdual}. Consider the following
{\em gedanken} calculation in the strongly coupled IIA string theory:\footnote{This
is only a {\em gedanken} calculation at the present time because the strongly coupled 
IIA string theory is, more precisely, M theory compactified 
on an $S^1$$\times$$S^1/{\rm Z}_2$, and we do not know how to calculate
in that 
theory beyond its low energy 11-dimensional supergravity limit.} we 
evaluate the pair correlation function of a pair of spatially
separated Wilson loops wrapping the eleventh dimension. The
limit of pointlike loops, and large spatial separation, corresponds 
to taking $R_{10}$ $\to$ $0$, and hence we recover a
result in {\em ten-dimensional} type IIA supergravity. Our thought experiment must,
therefore, have a precise analog in the factorization limit of a suitable 
weakly coupled perturbative type IIA string 
amplitude with the worldsheet topology of an annulus.
The relevant computation 
is as follows: consider the Polyakov path integral summing 
over worldsurfaces with the topology of an annulus, and with Dirichlet
boundary conditions on all ten embedding target space dimensions. 
We will require further that the boundaries are 
mapped to a pair of given pointlike loops, ${\cal C}_i$, ${\cal C}_f$, in
the target spacetime. The loops are taken to be 
spatially separated by a distance $R$ in the $X^8$ direction, lying within
the worldvolume of a stack of 32 coincident D8branes on a single $O8$ 
plane.  This augmented boundary value problem for embedded
Riemann surfaces with the topology of an annulus 
is the precise supersymmetric type IIA analog of a computation
carried out in 1986 by Cohen, Moore, Nelson, and Polchinski \cite{cmnp}
for the bosonic string theory: the off-shell closed string tree propagator 
between pointlike loops. An analysis of the type IIA macroscopic loop amplitude appears in a 
paper by myself and Novak in \cite{ferm}, and in generic background two-form
field. The factorization limit of the
 amplitude with pointlike loops 
yields the long-range
interaction of a pair of supergravity sources, and the result was found 
by me to be consistent with the tension of a Dirichlet (-2)brane \cite{flux}. 

\vskip 0.1in The detailed derivation of our expression for the
macroscopic loop amplitude is given in the Appendix. We perform
the sum over worldsheets with the topology of an annulus, and with
boundaries mapped to a pair of pointlike loops within the D8brane
worldvolume which also carries a constant two-form background.\footnote{One 
could, with no loss of generality, refer
to these as Wilson loops, as we have done in
\cite{wils,ferm,flux}, since the D8brane worldvolume carries
nonabelian gauge fields and the endpoint of open strings
transforms in the fundamental representation of the Yang-Mills
gauge group. In this section, our focus is on pointlike loops and
hence on the type IIA supergravity interpretation.} We will
orient the parallel loops, ${\cal C}_i$, ${\cal C}_f$, to lie in a
plane perpendicular to their distance of nearest separation, $R$,
along the spatial coordinate $X^{8}$. We emphasize that all ten
translation invariances of the pointlike loops are forbidden: we
have imposed the Dirichlet boundary conditions on all of the
$X^{\mu}$. Such boundary conditions can be implemented from first
principles on an extension of the covariant Polyakov string path
integral \cite{cmnp,wils,ferm}. The Polyakov action contributes a
classical piece corresponding to the saddle-point of the quantum
string path integral; the saddle-point is determined by the
minimum action worldsurface spanning the given loops ${\cal C}_i$,
${\cal C}_f$. The result for a generic classical solution of the
Polyakov action was given by Cohen, Moore, Nelson, and Polchinski
in Ref.\ \cite{cmnp}.

\vskip 0.1in The basic methodology for the computation of the
macroscopic loop amplitude is reviewed in detail in the Appendix.
We emphasize that worldsheet supersymmetry of the type IIB string
theory does not introduce any new features in the treatment of the
measure for moduli; this was clarified by myself and Novak in
\cite{ferm}. The main difference from Polchinski's analysis of the
exchange amplitude for the remaining members of the Dpbrane
spectrum, is implementation of the macroscopic loop constraints:
we wish to sum over all surfaces with the topology of an annulus,
but with boundaries mapped on to a pair of prescribed curves,
${\cal C}_i$, ${\cal C}_f$ in the embedding target spacetime. This
necessitates that all 10 worldsheet scalars in the path integral
are Dirichlet: the Dinstanton boundary condition of \cite{dbrane}.
But, in addition, one must preserve the full super-Weyl $\times$
superdiffeomorphism invariance in summing over maps from the
boundaries of the annulus to a given pair of loops in the target
spacetime. This aspect of the Wilson loop boundary value problem
was first addressed for pointlike loops in \cite{cmnp}, and
extended to the case of macroscopic loops in
\cite{cmnp,wils,ferm}. The result is an additional contribution to
the bosonic measure for moduli in the path integral coming from
the sum over boundary einbeins.

\vskip 0.1in Our result for the connected sum over worldsurfaces
with the topology of an annulus, and with boundaries mapped onto
spatially separated macroscopic loops, ${\cal C}_i$, ${\cal C}_f$,
of common length $L$ takes the form \cite{cmnp,ferm}:
\begin{eqnarray}
 {\cal A} &&= i \left [ L^{-1}(4\pi^2 \alpha^{\prime})^{1/2}
 \right ] \int_0^{\infty} {{dt}\over{2t}} \cdot (2t)^{1/2}
    \cdot e^{-R^2 t /2\pi\alpha^{\prime}}
 {{1 }\over{\eta(it)^8 }} \times
\nonumber\\
&&\quad \left [ \left ({{ \Theta_{00} (0 , it)}\over{ \eta (it) }}
\right )^4 - \left ( {{ \Theta_{0 1} (0 , it)}\over{ \eta (it) }}
\right )^4 -
 \left ( {{ \Theta_{10} (0 , it)}\over{ \eta (it) }} \right )^4 \right ].
 \label{eq:pdbre}
\end{eqnarray}
This result is derived in pedagogical detail in the Appendix. The
only change in the measure for moduli is the additional factor of
$(2t)^{1/2}$ contributed by the functional determinant of the
bosonic vector Laplacian evaluated on the one-dimensional boundary
\cite{cmnp,wils}. Perhaps not surprisingly, there are no
additional contributions in the supersymmetric type II amplitude
as was clarified by us in \cite{ferm}.

\vskip 0.1in The pre-factor in square brackets is of interest. A
priori, since we have broken translational invariance in all $10$
directions of the embedding worldvolume, we expect to see no
spacetime volume dependence in the prefactor. If we were only
interested in the point-like off-shell closed string propagator,
as in \cite{cmnp}, the result for the amplitude would be correct
without any need for a pre-factor. However, we have {\em required}
that the boundaries of the annulus are mapped to loops in the
embedding spacetime of an, a priori, fixed length $L$. Since a
translation of the boundaries in the direction of spacetime
parallel to the loops is equivalent to a boundary diffeomorphism,
we must divide by the (dimensionless) factor: $L (4\pi^2
\alpha^{\prime})^{-1/2}$. This accounts for the volume dependence
in the pre-factor present in our final result. Note that for more
complicated loop geometries, the normalization of this expression
can take a more complicated form \cite{wils}. Further discussion
appears in Appendix A.2.

\vskip 0.1in Let us take the factorization limit of the amplitude,
expressing the Jacobi theta functions in an expansion in powers of
$q$$=$$e^{-2\pi/t}$. The small $t$ limit is dominated by the
lowest-lying closed string modes and the result is:
\begin{eqnarray}
 {\cal A} =&& i \left [ L^{-1}(4\pi^2 \alpha^{\prime})^{1/2}
 \right ]
   \int_0^{\infty} dt\cdot (2t)^{-1/2} \cdot t^{4} \cdot q^{-1}
    \left ( 1 + 8 q + O(q^2) \right ) e^{-R^2 t
    /2\pi\alpha^{\prime}}\cr
\to&& i L^{-1}
 8 \cdot 2^{-8}(4\pi^2 \alpha^{\prime})^{5} \pi^{-9/2} \Gamma \left ( {{9}\over{2}} \right )
 |R|^{-9} \quad .
\label{eq:resulttafbt}
\end{eqnarray}
Repeating the steps in Polchinski's calculation of the Dbrane
tension \cite{dbrane,polbook,zeta}, we infer the existence of a
Dirichlet (-2)brane in the massive type II string theory with
tension:
\begin{equation}
\tau_{-2}^2 = {{\pi}\over{\kappa^2}} (4\pi^2 \alpha^{\prime} )^{5}
\quad . \label{eq:tension-2}
\end{equation}

\vskip 0.1in One of the interesting properties of the BPS states
in the type II brane spectrum is their isomorphic
ultraviolet-infrared behaviour: the force law between Dpbranes
takes identical form at short and long distances. The pointlike source
of zeroform field strength 
is no different in this respect, as already noted in \cite{flux}.
Let us quote the result for the full short-distance asymptotic
expansion of the force between these sources in the presence of a
constant electromagnetic background field. We comment that this
formal expression was derived in \cite{wils,ferm}, prior to a full
appreciation of its physical interpretation as the short distance
force law for point sources of the IIA scalar field strength described by 
us in
\cite{flux}:
\begin{eqnarray}
V(r,\alpha) =&& - L^{-1} (4\pi^2 \alpha^{\prime})^{1/2}
   {{{\rm tanh} ( \pi \alpha) }\over{ \pi \alpha  }} {{1}\over{r}}
\nonumber\\
&&\quad \times
    \left [ \sum_{k=1}^{\infty} C_k z^{2k} \gamma \left ( 2k+\half , 1/z \right )
+ \sum_{k=1}^{\infty} \sum_{m=1}^{\infty} C_{k,m} z^{2(k+m)}
\gamma \left ( 2 (k+m) + \half ,
 1/z \right ) \right ] ,
\label{eq:potstas}
\end{eqnarray}
where the $\gamma(2n + \half ,1/z)$ are the incomplete gamma
functions. The asymptotic expansion has been developed in terms of
the natural choice of dimensionless variable: $z$ $=$ $r_{\rm
min}^2/r^2$. Here, $r_{\rm min}^2$ $=$ $2\pi \alpha^{\prime} (\pi
\alpha)$ $=$ $4\pi^2 \alpha^{\prime} |\alpha^{\prime}
{\bar{F}}_2|$, is the minimum distance accessible in open string
theory in the presence of a constant background electromagnetic
field, $|{\bar{F}}_2|$ \cite{dkps}. The coefficients in the
asymptotic expansion are given by \cite{wils}:
\begin{eqnarray}
C_k =&& {{4(-1)^k (2^{2k} - 4)}\over{(2k!)}} \pi^{2k}
\nonumber\\
C_{k,m} =&& {{8 (-1)^k (2^{2m-1} - 1)}\over{(2k!)(2m!) }} |B_{2m}|
(2^{2k} - 4)\pi^{2(k+m)} \quad ,
\nonumber\\
\label{eq:potstasc}
\end{eqnarray}
where the $B_{2m}$ are the Bernoulli numbers. Notice that the
coefficient of the $k$$=$$1$ term in both sums vanishes, so that
the leading contribution to the short distance potential behaves
as $-\alpha^4/r^9$, precisely as expected for point sources of the
zeroform field strength \cite{polbook}:
\begin{eqnarray}
V(r,\alpha) =&& - L^{-1} (4\pi^2 \alpha^{\prime})^{1/2}
\int_0^{\alpha^{-1}} dt e^{-r^2t/2\pi\alpha^{\prime}}
  (2t)^{1/2} {{{\rm tanh}  \pi \alpha }\over{{\rm Sin} \pi \alpha t }}
    \left [ 12 + 4 {\rm Cos} (2\pi \alpha t) - 16 {\rm Cos} (\pi \alpha t) \right ]
\nonumber\\
=&& - L^{-1} {{\alpha^4}\over{r^9}} 2^6 \pi^{11/2} \Gamma \left (
{{9}\over{2}} \right ) \alpha^{\prime 9/2} + O(\alpha^6) \quad .
\label{eq:potst}
\end{eqnarray}
The range of integration has been taken to span the first period
of the sine function, $0$ $\le$ $\alpha t$ $\le $ $1$, as
discussed in \cite{polbook,wils}. This result is a 
concrete indication of pointlike sources of the supergravity 
zeroform field strength, 
an identification first made by us
in \cite{flux}.

\section{Conclusions}

\vskip 0.1in We have found a satisfying resolution to the puzzle
raised in the Introduction that appears to soldifies the case in
favor of an {\em eleven}-dimensional electric-magnetic duality in
String/M Theory. Poincare-Hodge duality in 11 dimensions would
imply a Dirichlet pbrane spectrum with twelve elements, with $p$
taking values in the range: $-2$ $\le$ $p$ $\le$ $9$. This is
precisely the spectrum of Dpbrane charges for which we have found
evidence in the worldsheet formalism of the perturbative type I
and type II supersymmetric string theories.

\vskip 0.1in In particular, we have demonstrated explicitly that
Polchinski's worldsheet calculation of the Dpbrane tension, with
$p$ taking values in the range $-1$ $\le$ $p$ $\le$ $9$ in the
type II string theories \cite{dbrane}, admits a unique extension
that is natural both from the perspective of the boundary value
problem for Riemann surfaces, as well as from the target spacetime
perspective of generalized electric-magnetic duality in the R-R
soliton spectrum. We have evaluated the pair correlation 
function of a pair of spatially
separated Wilson loops wrapping the eleventh dimension. The
limit of pointlike loops, and large spatial separation, corresponds 
to taking $R_{10}$ $\to$ $0$, and it must have a precise analog 
in the factorization limit of a suitable 
weakly coupled perturbative type IIA string 
amplitude with the worldsheet topology of an annulus.
We have shown that the requisite worldsheet calculation yields
the tension of a pointlike source of the zeroform field strength
in String/M theory.

\vskip 0.1in The
 massive soliton strings in the type IIA 
D0-F1-D8 system \cite{jmo,troost} 
had the
target spacetime geometry of a solitonic F1 string with
D0brane charge spread smoothly along the length of an F1 string
extending in the direction orthogonal to the worldvolume of an
D8brane. It is natural to suspect that such a type I$^{\prime}$
soliton string can be uplifted to a soliton background for M
theory compactified on $S^1$$\times$$S^1/{\rm Z}_2$ with
the following target space geometry: a fundamental string
extending along the interval $X^{10}$ between orientifold planes
in eleven dimensions, terminating in the D9branes on the \lq\lq
walls" bounding the interval. It would be most interesting if the
arguments in this paper, and the supergravity results in
\cite{jmo,troost}, could be extended to show that the endpoint of
the F1 string behaves like a source of Dirichlet
(-2)brane charge in eleven dimensions. The zero form, and eleven form, field strengths
are Poincare duals in eleven dimensions, and we would like to
interpret the absence of a D(-3)brane in the Dpbrane spectrum of
the type II string theories as unambiguous indication of {\em the 11d 
nature of electric
magnetic duality in the Dbrane spectrum of the type II string theories}.

\vskip 0.1in So what does this tell us more generally about
Poincare-Hodge duality in String/M theory? In 
\cite{mtheory}, we have explained why Poincare duality in eleven
dimensions, consistent with the appearance of precisely $12$
elements in the Dirichlet pbrane spectrum of the type II string 
theories, $-2$ $\le$ $p$ $\le$
$9$, could be perfectly compatible with 
a fundamental theory of emergent spacetime based upon degrees of
freedom that are {\em zero-dimensional} U(N) matrices \cite{mtheory}.
The emergence of dualities in the large $N$ limits of this theory has to do with the
choice of Duality group, determining the precise form of the matrix
Lagrangian \cite{mtheory}.
Notice that as many as eleven noncompact coordinates
can indeed emerge in the different large $N$ limits of the matrix 
Lagrangians described in \cite{mtheory}. The full significance of 
electric-magnetic duality in nonperturbative 
string/M theory remains to be discovered.\footnote{An 
expanded discussion of electric-magnetic dualities from this perspective
appears in \cite{landscape}.}

\vskip 0.2in \centerline{\bf ACKNOWLEDGEMENTS}

\medskip This paper was begun in the stimulating
environs of the Aspen Center for Physics, and I would like to
acknowledge their hospitality in Summer 2004. I thank 
Hermann Nicolai, John Schwarz, and 
Axel Kleinschmidt for interesting remarks on 11d electric-magnetic
duality. The worldsheet calculations, completed in partial collaboration 
with Eric Novak, date to Spring 2000. 

\vskip 0.4in \noindent{\Large\bf Appendix: Macroscopic Loop
Amplitudes}

\vskip 0.1in For completeness, we are presenting this derivation
in full pedagogical detail. The macroscopic loop amplitude sums
over worldsurfaces with two boundaries. Thus, there is only one
contributing worldsheet topology at leading order in string
perturbation theory, even in the unoriented open and closed string
theories.

\vskip 0.2in \noindent{\large\bf A.1  Review of the Dpbrane
Exchange Amplitude}

\vskip 0.1in We begin by reviewing the computation of the one-loop
amplitude of bosonic open and closed string theory from
world-surfaces with the topology of a cylinder following
Polchinski's analogous first principles path integral derivation
for the torus in \cite{poltorus}, and including the extensions to
$p$$+$$1$ Dirichlet boundaries \cite{dbrane}, with $-1$ $\le$ $p$
$\le$ $9$, and an external two form field strength
\cite{bachas,call,nal}. From the perspective of the closed string
channel, this amplitude represents the tree-level propagation of a
single closed string, exchange between a spatially-separated pair
of Dpbranes. A crucial observation is as follows: although the
Dpbrane vacuum corresponds to a spontaneous breaking of
translation invariance in the bulk $25$$-$$p$ dimensional space
orthogonal to the pair of Dpbranes, notice that spacetime
translational invariance is preserved within the
$p$$+$$1$-dimensional worldvolume of each Dpbrane. At the end of
this sub-section, we can specialize to the case of Dinstantons
with all $26$ embedding target space scalars satisfying Dirichlet
boundary conditions.

\vskip 0.1in One further clarification is required to distinguish
results in the presence, or absence, of Dirichlet p-branes. In the
absence of Dpbranes, the boundary of the worldsheet can lie in all
$26$ dimensions, and we impose Neumann boundary conditions on all
$d$$=$$26$ scalars. This gives the traditional open and closed
string theory, whose supersymmetric generalization is the type IB
string theory. T-dualizing $25$$-$$p$ embedding coordinates gives
the open and closed string theory in the background geometry of a
pair of Dpbranes \cite{dbrane}. Its supersymmetric generalization
is the type I$^{\prime}$ string theory, when $p$ is even, and the
type IB string theory in generic Dpbrane background, when $p$ is
odd \cite{dbrane}. The Dpbranes define the hypersurfaces bounding
the compact bulk spacetime, which is $(25$$-$$p)$-dimensional.
Since the bulk spacetime has edges, these $25$$-$$p$ embedding
coordinates are Dirichlet worldsheet scalars. It is conventional
to align the Dpbranes so that the distance of nearest separation,
$R$, corresponds to one of the Dirichlet coordinates, call it
$X^{25}$.

\vskip 0.1in In the presence of a pair of Dpbranes, the classical
worldsheet action contributes a background term given by the
Polyakov action for a string of length $R$ stretched between the
Dpbranes \cite{dbrane,polbook,zeta}:
\begin{equation}
S_{\rm cl} [G,g] = \quaft \int d^2 \sigma {\sqrt{g}} g^{ab} ~
G^{25,25} (X)
   \partial_a X^{25} \partial_b X^{25}
 =  {{1}\over{2\pi \alpha^{\prime}}} R^2 t \quad ,
\label{eq:backg}
\end{equation}
where the second equality holds in the critical dimension on open
world-surfaces of vanishing Euler number. The background
dependence, $e^{-S_{\rm cl}[G,g]}$, appears in the sum over
connected open Riemann surfaces of any topology, orientable or
nonorientable. Notice that the background action is determined
both by the fiducial worldsheet metric, $g_{ab}$, {\em and} by the bulk
spacetime metric, $G_{\mu\nu}[X]$. Notice also that the boundary
of an open worldsheet is now required to lie within the
worldvolume of the Dpbrane, although the worldsheet itself is
embedded in all $26$ spacetime dimensions.

\vskip 0.1in The metric on the generic annulus can be
parameterized by a single real worldsheet modulus, $t$, and it
takes the form:
\begin{equation}
ds^2 = e^{\phi} ( (d\sigma^1)^2 + 4t^2 (d \sigma^2)^2 ), \quad
{\sqrt{g}} = e^{\phi} 2t , \quad 0 \le t \le \infty \quad ,
\label{eq:metrica}
\end{equation}
with worldsheet coordinates, $\sigma^a$, $a$$=$$1,2$,
parameterizing a square domain of unit length. $2t$ is the
physical length of either boundary of the annulus as measured
in the two-dimensional field theory on the worldsheet. The
differential operator mapping worldsheet vectors, $\delta
\sigma^a$, to symmetric traceless tensors, usually denoted $(P_1
\delta \sigma)_{ab}$, has only one zero mode on the annulus. This
is the constant diffeomorphism in the direction tangential to the
boundary: $\delta \sigma^2_0$. Likewise, the analysis of the zero
modes of the scalar Laplacian must take into account the Dpbrane
geometry: the $p+1$ noncompact embedding coordinates satisfying
Neumann boundary conditions are treated exactly as in the case of
the torus. The $25-p$ Dirichlet coordinates lack a zero mode.
Thus:
\begin{eqnarray}
1 =&& \int d\delta X  e^{-\half |\delta X|^2 } =
 \prod_{\mu=0}^{p} \int d\delta {\bar{x}}
e^{-\half (\delta {\bar{x}}^{\mu})^2 \int d^2 \sigma {\sqrt{g}} }
 \int d\delta X^{\prime} e^{-\half |\delta X^{\prime} |^2 } \cr
=&&
 (2\pi )^{(p+1)/2} \left ( \int d^2 \sigma {\sqrt{g}} \right )^{-(p+1)/2}
 \int d\delta X^{\prime} e^{-|\delta X^{\prime} |^2/2 }
 \quad .
\label{eq:gaussxa}
\end{eqnarray}
The analysis of the diffeomorphism and Weyl invariant measure for
moduli follows precisely as for the torus, differing only in the
final result for the Jacobian \cite{zeta}. We have:
\begin{eqnarray}
1 =&& \int d g d X e^{-\half |\delta g|^2 - \half |\delta X|^2 }
\cr =&& \left ( {\rm det}  Q_{22} \right )^{-1/2}
 {{ \left ( \int d^2 \sigma {\sqrt{g}} \right )^{1/2+(p+1)/2} }\over{
     (2\pi)^{( p+1)/2} }}
\left ({\rm det}^{\prime} {\cal M} \right )^{1/2}  \int
 (d\phi d\delta \sigma)^{\prime} d t  dX^{\prime}
           e^{-\half |\delta g|^2 - \half |\delta X|^2 } ,
\label{eq:jacoa}
\end{eqnarray}
where $({\rm det} Q_{22})$ $=$ $2t$ in the critical dimension,
cancelling the factor of $2t$ arising from the normalization of
the integral over the single real modulus. As shown in
\cite{zeta}, ${\cal M}$ takes the form:
\begin{equation}
\left ( {\rm det}^{\prime} {\cal M} \right )^{1/2} = \left ({\rm
det}^{\prime} ~ 2 ~ \Delta^c_d \right )^{1/2} \left (
{{1}\over{2t}} \right ) = {{ 1}\over{2}} (2t)^{-1} {\rm
det}^{\prime} \Delta = \half (2t)^{-1}
\prod_{n_2=-\infty}^{\infty} \prod_{n_1=-\infty}^{\infty \prime }
\omega_{n_2 , n_1} \quad . \label{eq:vecta}
\end{equation}
The Laplacian acting on free scalars on an annulus with boundary
length $2t$ takes the form, $\Delta$$=$$(2t)^{-2}\partial_2^2
$$+$$ \partial_1^2$, with eigenspectrum:
\begin{equation}
\omega_{n_2 , n_1 } = {{\pi^2}\over{t^2}} ( n_2^2 + n_1^2 t^2 ) ,
\quad \Psi_{n_2,n_1} = {{1}\over{{\sqrt{2t}}}}
 e^{2\pi i n_2 \sigma^2}  {\rm Sin} (\pi n_1 \sigma^1 )
\quad , \label{eq:deegtermaf}
\end{equation}
where the subscripts take values in the range
$-\infty$$\le$$n_2$$\le$$\infty$, and $n_1$$\ge$$0$ for a Neumann
scalar, or $n_1$$\ge$$1$ for a Dirichlet scalar.

\vskip 0.1in In the case of a background electromagnetic field,
${\cal F}_{p-1,p}$, it is convenient to complexify the
corresponding pair of scalars: $Z$$=$$X^p $$+$$i X^{p-1}$
\cite{call,bachas,zeta}. They satisfy the following twisted
boundary conditions:
\begin{eqnarray}
\partial_1 {\rm Re} Z =&& {\rm Im} Z = 0 \quad \sigma^1 = 0
\\
\partial_1 {\rm Re} e^{-i\phi} Z =&& {\rm Im} e^{-i\phi} Z = 0 \quad \sigma^1 = 1
\label{eq:bc}
\end{eqnarray}
Expanding in a complete set of orthonormal eigenfunctions gives:
\begin{equation}
Z = \sum_{n_2, n_1} z_{n_2, n_1} \Psi_{n_2 , n_1 } =
{{1}\over{{\sqrt{2t}}}}
 e^{2\pi i n_2 \sigma^2}  {\rm Sin} \pi (n_1 + \alpha )\sigma^1
\quad , \label{eq:set}
\end{equation}
where $\pi\alpha$$=$$\phi$, and $\pi$$-\phi$, respectively
\cite{zeta}, and subscripts take values in the range
$-\infty$$\le$$n_2$$\le$$\infty$, $n_1$$\ge$$0$. The twisted
complex scalar has a discrete eigenvalue spectrum on the annulus
given by:
\begin{equation}
\omega_{n_2 , n_1 } (\alpha) = {{\pi^2}\over{t^2}} ( n_2^2 +
(n_1+\alpha)^2 t^2 ) \quad . \label{eq:deegterma}
\end{equation}
Thus, the connected sum over worldsurfaces with the
topology of an annulus embedded in the target spacetime geometry of a
pair of parallel Dpbranes separated by a distance $R$ in the
direction $X^9$, and in the absence of a magnetic field, takes the
form \cite{zeta}:
\begin{equation}
 W_{\rm ann} =  \prod_{\mu = 0}^p L^{\mu}
   \int_0^{\infty} {{dt}\over{2t}}
   (8\pi^2 \alpha^{\prime}t)^{-(p+1)/2} \eta (it )^{-24}
                       e^{-R^2 t /2\pi\alpha^{\prime}}
\quad . \label{eq:resulttaf}
\end{equation}
In the presence of a worldvolume electromagnetic field, ${\cal
F}_{p-1,p}$, the scalars $X^{p-1}$, $X^{p}$, are complexified.
Substituting the result for the eigenspectrum of the twisted
complex scalar gives:
\begin{equation}
 W_{\rm ann}(\alpha) = \prod_{\mu = 0}^{p-2} L^{\mu}
   \int_0^{\infty} {{dt}\over{2t}}
   (8\pi^2 \alpha^{\prime}t)^{-(p-1)/2} \eta (it )^{-22}
                       e^{-R^2 t /2\pi\alpha^{\prime}}
 {{ e^{\pi t \alpha^2} \eta(it)}\over{ i \Theta_{11} (it \alpha ,it) }}
\quad , \label{eq:resultta}
\end{equation}
with $\alpha$$=$$i\phi/\pi$ and $q$$=$$e^{-2\pi t}$.

\vskip 0.2in \noindent{\large\bf A.2  The Extension to Macroscopic
Boundary Loops}

\vskip 0.1in We move on to the extension to macroscopic boundary
loops, first explored by Cohen, Moore, Nelson, and Polchinski
\cite{cmnp}. So far, we have performed the Weyl $\times$
Diffeormorphism invariant sum over all worldsurfaces with the
topology of an annulus, and with fully Dirichlet boundaries. We
now wish to perform a gauge invariant sum over all maps of the
boundaries of the annulus to a pair of fixed curves, ${\cal C}_i$,
${\cal C}_f$ in the embedding target spacetime. The main
modification is to the sum over metrics on the boundary: it would
be too strong to impose Dirichlet boundary condition on the
deformations of the worldsheet metric. Instead, we can parametrize
the generic map from worldsheet boundary to target space loop in
terms of a boundary einbein \cite{cmnp}. What remains is to
perform the sum over the set of {\em gauge-inequivalent} maps of 
worldsheet boundary to target space loop.

\vskip 0.1in We should pause to point out that it is this last
step which was not completed in \cite{cmnp}, except for the case
of pointlike loops, where it is trivial. In \cite{wils}, we
realized that, for large loop length, the gauge invariant sum over
arbitrary deformations of the einbein simply amounts to adapting
the sum over gauge-inequivalent deformations of the bulk
worldsheet metric to those of the boundary einbein
\cite{polyakov,poltorus}. For more topologically nontrivial Wilson
loop configurations, the normalization of the macroscopic loop
amplitude can indeed contain additional numerical factors arising
from the discrete component of the sum over inequivalent maps. But
the worldsheet modulus dependence in the measure of the modular
 integral will be
unchanged from the simplest case of 
single-winding circular Wilson loops. The modulus dependence in
the measure of the string path integral is of crucial physical
significance, since it determines the precise power law behavior of the
long-range potentials obtained in the low energy field theory limits 
of the string amplitude.
 See the 
comments below Eq.\ (30), and in Footnote 6.

\vskip 0.1in For simplicity, it is convenient to align the
macroscopic loops, ${\cal C}_i$, ${\cal C}_f$, which we choose to
have the common length $L$, such that their distance of nearest
separation, $R$, is parallel to a spatial coordinate, call it
$X^{25}$. As in the previous sub-section, the Polyakov action
contributes a classical piece corresponding to the saddle-point of
the quantum path integral: the saddle-point is determined by the
minimum action worldsurface spanning the given loops ${\cal C}_i$,
${\cal C}_f$. The result for a generic classical solution of the
Polyakov action was derived in \cite{cmnp}. For coaxial circular
loops in a flat spacetime geometry, the result is identical to
that for a spatially separated pair of generic Dpbranes in flat
spacetime, namely Eq.\ (\ref{eq:backg}). Notice, in particular,
that there is no $L$ dependence in the saddle-point action as a
consequence of the Dirichlet boundary condition on all $26$
scalars. As in the case of Dinstanton boundary conditions
considered in \cite{dbrane}, we evaluate the determinant of the
scalar Laplacian for all $26$ embedding coordinates with the
Dirichlet boundary condition. Notice that there is no contribution
from coordinate zero modes, since all of the $X^{\mu}$ are
Dirichlet. Thus, the usual box-regularized spacetime volume
dependence originating in the Neumann sector is absent, precisely
as in the vacuum of a pair of Dinstantons. The analog of Eq.\
(\ref{eq:gaussxa}) reads:
\begin{equation}
1 = \int d\delta X  e^{-\half |\delta X|^2 } =
 \int d\delta X^{\prime} e^{-|\delta X^{\prime} |^2/2 }
 \quad .
\label{eq:gaussxam}
\end{equation}

\vskip 0.1in The crucial difference in the path integral
computation when the boundaries of the annulus are mapped to {\em
macroscopic} loops in embedding spacetime has to do with the
implementation of boundary reparameterization invariance: we must
include in the path integral a sum over all possible maps of the
worldsheet boundary to the loops ${\cal C}_i$, ${\cal C}_f$
\cite{cmnp}. Notice that the analysis of reparametrization
invariance in the bulk of the worldsheet is unaltered. As a
consequence, the conditions for Weyl invariance, and for the
crucial decoupling of the Liouville mode, are unchanged.

\vskip 0.1in Let us proceed with the analysis of the measure
following the steps in \cite{polyakov,poltorus}. The differential
operator mapping worldsheet vectors, $\delta \sigma^a$, to
symmetric traceless tensors, usually denoted $(P_1 \delta
\sigma)_{ab}$, has only one zero mode on the annulus. This is the
constant diffeomorphism in the direction tangential to the
boundary: $\delta \sigma^2_0$. The analysis of the diffeomorphism
and Weyl invariant measure for moduli follows precisely as for the
annulus.\footnote{A pedagogical derivation can be found in the
electronic review \cite{zeta}}. The only difference is an
additional contribution from the vector Laplacian, accounting for
diffeomorphisms of the metric which are nontrivial on the boundary
\cite{cmnp}. The analog of Eq.\ (\ref{eq:jacoa}) now takes the
form:
\begin{eqnarray}
1 =&& \int d e \int d g d X e^{-\half |\delta g|^2 - \half |\delta
X|^2 - \half |\delta e|^2} \cr =&& \left ( {\rm det}  Q_{22}
\right )^{-1/2}
  \int d^2 \sigma {\sqrt{g}}
\left ({\rm det}^{\prime} {\cal J} \right )^{1/2} \left ({\rm
det}^{\prime} {\cal M} \right )^{1/2}  \int
 (d\phi d\delta \sigma)^{\prime} d t  dX^{\prime}
           e^{-\half |\delta g|^2 - \half |\delta X|^2 - \half |\delta e|^2 }
           ,\cr
           &&
\label{eq:jacoab}
\end{eqnarray}
where $({\rm det} Q_{22})$ $=$ $2t$ in the critical dimension,
canceling the factor of $2t$ arising from the normalization of
the integral over the single real modulus. The functional
determinant of the vector Laplacian acting in the worldsheet bulk
takes the form:
\begin{equation}
\left ( {\rm det}^{\prime} {\cal M} \right )^{1/2} = \left ({\rm
det}^{\prime} ~ 2 ~ \Delta^c_d \right )^{1/2} \left (
{{1}\over{2t}} \right ) = {{ 1}\over{2}} (2t)^{-1} {\rm
det}^{\prime} \Delta = \half (2t)^{-1}
\prod_{n_2=-\infty}^{\infty} \prod_{n_1=-\infty}^{\infty \prime }
\omega_{n_2 , n_1} \quad , \label{eq:vectab}
\end{equation}
and the infinite product is computed as in appendix A.3. The
functional determinant of the operator ${\cal J}$ can likewise be
expressed in terms of the functional determinant of the Laplacian
acting on free scalars on the one-dimensional boundary,
parameterized here by $\sigma^2$ \cite{cmnp}. Thus, for boundary
length $2t$, we have $\Delta_b$$=$$(2t)^{-2}\partial_2^2 $, with
eigenspectrum:
\begin{equation}
\omega_{n_2 } = {{\pi^2}\over{t^2}} n_2^2  , \quad \Psi_{n_2} =
{{1}\over{{\sqrt{2t}}}}
 e^{2\pi i n_2 \sigma^2}
\quad , \label{eq:deegtermafb}
\end{equation}
where the subscripts take values in the range
$-\infty$$\le$$n_2$$\le$$\infty$. The result can be read off as a
special case of the expressions in appendix A.3.

\vskip 0.1in Thus, the connected sum over worldsurfaces with the
topology of an annulus with boundaries mapped onto spatially
separated macroscopic loops, ${\cal C}_i$, ${\cal C}_f$, of common
length $L$ takes the form \cite{cmnp,wils}:
\begin{equation}
 {\cal A}_{i,f} = \left [ L^{-1}(4\pi^2 \alpha^{\prime})^{1/2}
 \right ]
   \int_0^{\infty} {{dt}\over{2t}} \cdot (2t)^{1/2} \cdot
    \eta (it )^{-24} e^{-R^2 t /2\pi\alpha^{\prime}}
\quad . \label{eq:resulttafbs}
\end{equation}
The only change in the measure for moduli is the additional factor
of $(2t)^{1/2}$ contributed by the functional determinant of
${\cal J}$. The pre-factor in square brackets is of interest;
recall that there is no spacetime volume dependence in this
amplitude since we have broken translational invariance in all
$26$ directions of the embedding spacetime. If we were only
interested in the point-like off-shell closed string propagator,
as in \cite{cmnp}, the result as derived is correct without any
need for a pre-factor. However, we have {\em required} that the
boundaries of the annulus are mapped to loops in the embedding
spacetime of an, a priori, fixed length $L$. Since a translation
of the boundaries in the direction of spacetime parallel to the
loops is equivalent to a boundary diffeomorphism, we must divide
by the (dimensionless) factor: $L (4\pi^2
\alpha^{\prime})^{-1/2}$. This accounts for the pre-factor present
in our final result. Note that for more complicated loop
geometries, including the possibility of Wilson loops with multiple windings,
or even cusps and corners \cite{wils}, the numerical 
pre-factor in this expression can take a far more complicated form.

\vskip 0.1in \noindent{\large\bf A3 The Macroscopic Loop Amplitude in
Type II String Theories}

\vskip 0.1in The derivation of the gauge invariant measure for
moduli given for the bosonic string can be easily extended to the
case of the supersymmetric unoriented open and closed string
theories. in generic type II string backgrounds with Dbranes
\cite{dbrane,polbook,nal,ferm,flux}. We begin with the
contribution from worldsurfaces with the topology of an annulus in
the presence of a background electromagnetic field, and in the
background spacetime geometry of a pair of Dpbranes separated by a
distance $R$. This expression was derived in \cite{wils}:
\begin{eqnarray}
 W_{ann-I} (\alpha) =&&  \prod_{\mu=0}^{p-2}
   \int_0^{\infty} {{dt}\over{2t}} (8\pi^2 \alpha^{\prime}t)^{-(p-1)/2}
\eta (it )^{-6} e^{-R^2 t /2\pi\alpha^{\prime}}
  {{ e^{\pi t \alpha^2} \eta(it)}\over{ i \Theta_{11} (it \alpha ,it) }}
\nonumber\\
\quad && \quad \times \prod_{n_2=0}^{\infty}
\prod_{n_1=-\infty}^{\infty \prime} \left (  {\rm det}_{\rm ann-I}
\Delta_{n_2 + \half , n_1 + \half + \alpha } \right )^{1} \left (
{\rm det}_{\rm ann-I} \Delta_{n_2 + \half , n_1 + \half } \right
)^{3} , \label{eq:rannulusf}
\end{eqnarray}
where we have included the contribution from worldsheet bosonic
fields derived in the previous section.

\vskip 0.1in Let us understand the eigenvalue spectrum of the
worldsheet fermions in more detail. Recall that the functional
determinant of the two-dimensional Dirac operator acting on a pair
of Majorana Weyl fermions satisfying twisted boundary conditions
is equivalent, by Bose-Fermi equivalence, to the functional
determinant of the scalar Laplacian {\em raised to the inverse
power}. This provides the correct statistics. In addition, we have
the constraint of world-sheet supersymmetry. This requires that
the four complexified Weyl fermions satisfy identical boundary
conditions in each sector of the theory in the $\sigma^1$
direction. For a complex Weyl fermion satisfying the boundary
condition:
\begin{eqnarray}
\psi (1,\sigma^2) =&& - e^{\pi i a} \psi(0,\sigma^2)
\nonumber\\
\psi (\sigma^1 , 1 ) =&& - e^{ \pi i b} \psi(\sigma^1 , 0) \quad ,
\label{eq:ferbc}
\end{eqnarray}
the Bose-Fermi equivalent scalar eigenspace takes the form:
\begin{equation}
\Psi_{n_2 + \half , n_1 + \half (1+a)  } = {{1}\over{\sqrt{2t}}}
e^{ 2\pi i (n^2 + \half (1\pm b) )\sigma^2} {\rm Cos} \pi (n_1 +
\half (1 \pm a) ) \quad , \label{eq:fermseig}
\end{equation}
where we sum over $-\infty$$\le$$n_2$$\le$$\infty$, $n_1$$\ge$$0$.
Notice that the unrotated oscillators are, respectively,
half-integer or integer moded as expected for the scalar
equivalent of antiperiodic or periodic worldsheet fermions.
Finally, we must sum over periodic and antiperiodic sectors,
namely, with $a$, $b$ equal to $0$, $1$. As reviewed in the
appendix, weighting the $(a , b)$ sector of the path integral by
the factor $e^{  \pi i a b}$ gives the following result for the
fermionic partition function \cite{polbook}:
\begin{eqnarray}
Z^{a}_{b} (\alpha , q) =&& q^{\half a^2 - {{1}\over{24}}} e^{  \pi
i ab}
 \prod_{m=1}^{\infty} \left [ ( 1+ e^{  \pi i b } q^{m - \half(1 - a) + \alpha })
  ( 1+ e^{- \pi i b } q^{m - \half (1+ a)  - \alpha } )\right ]
\nonumber\\
\quad \equiv&&  {{1}\over{e^{\pi t \alpha^2} \eta(it) }} \Theta_{
a    b }  (\alpha it , it ) \quad . \label{eq:theta}
\end{eqnarray}
We have included a possible rotation by $\alpha$ or $1$$-$$\alpha$
as in the previous section. This applies for the Weyl fermion
partnering the twisted complex worldsheet scalar. Substituting in
the path integral, and summing over $a$, $b$$=$$0$, $1$, for all
fermions, and over $\alpha$ and $1$$-$$\alpha$ for the Weyl
fermion partnering the twisted complex scalar, gives the result:
\begin{eqnarray}
 W_{\rm ann-I} &&= \prod_{\mu=0}^{p-2} L^{\mu}
   \int_0^{\infty} {{dt}\over{2t}}
(8\pi^2 \alpha^{\prime}t)^{-(p-1)/2} \eta (it )^{-6}
      e^{-R^2 t /2\pi\alpha^{\prime}}
 \left [ {{ e^{\pi t \alpha^2} \eta(it)}\over{ \Theta_{11} (it \alpha ,it) }}
\right ]
\nonumber\\
\quad && \times \left [ {{ \Theta_{0 0 } (it \alpha , it)}\over{
e^{\pi t \alpha^2} \eta (it) }}
 \left ( {{ \Theta_{00} (0 , it)}\over{ \eta (it) }} \right )^3
-
 {{ \Theta_{0 1 } (it\alpha , it)}\over{ e^{\pi t \alpha^2} \eta (it) }}
 \left ( {{ \Theta_{0 1} (0 , it)}\over{ \eta (it) }} \right )^3
-
 {{ \Theta_{1 0} (it\alpha , it)}\over{ e^{\pi t \alpha^2} \eta (it) }}
 \left ( {{ \Theta_{10} (0 , it)}\over{ \eta (it) }} \right )^3
\right ]
\nonumber \\
\label{eq:rannulusfer}
\end{eqnarray}
where we have used the fact that $\Theta_{11 }(0,it)$ equals zero.
The analogous expression for the macroscopic loop amplitude can be
straightforwardly written down, since the only changes are in the
analysis of boundary deformations of the worldsheet metric. More
details can be found in \cite{ferm}. The result takes the form
\cite{ferm,flux}:
\begin{eqnarray}
 {\cal A} &&= i \left [ L^{-1}(4\pi^2 \alpha^{\prime})^{1/2}
 \right ] \int_0^{\infty} {{dt}\over{2t}} \cdot (2t)^{1/2}
    \cdot e^{-R^2 t /2\pi\alpha^{\prime}}
 \left [ {{ e^{\pi t \alpha^2} \eta(it)}\over{ \Theta_{11} (it \alpha ,it) }}
\right ]
\nonumber\\
\quad && \times \left [ {{ \Theta_{0 0 } (it \alpha , it)}\over{
e^{\pi t \alpha^2} \eta (it) }}
 \left ( {{ \Theta_{00} (0 , it)}\over{ \eta (it) }} \right )^3
-
 {{ \Theta_{0 1 } (it\alpha , it)}\over{ e^{\pi t \alpha^2} \eta (it) }}
 \left ( {{ \Theta_{0 1} (0 , it)}\over{ \eta (it) }} \right )^3
-
 {{ \Theta_{1 0} (it\alpha , it)}\over{ e^{\pi t \alpha^2} \eta (it) }}
 \left ( {{ \Theta_{10} (0 , it)}\over{ \eta (it) }} \right )^3
\right ]
\nonumber \\
 \label{eq:pdbree}
\end{eqnarray}

\vskip 0.2in \noindent{\large\bf A.4  Zeta-Regularized Eigenvalue
Spectrum of the Twisted Scalar}

\vskip 0.1in The regularization of a divergent sum over the
discrete eigenvalue spectrum of a self-adjoint differential
operator by the zeta function method can always be carried out in
closed form when the eigenvalues are known explicitly \cite{hawk}.
This is the case for all of the infinite sums encountered in
one-loop superstring amplitudes, even in the presence of generic 
background fields
\cite{poltorus,polbook}.\footnote{A
pedagogical review of Polchinski's calculation of the
zeta-regularized functional determinant of the scalar Laplacian on
the torus \cite{poltorus} can be found in the electronic review
\cite{zeta}, which also contains the analogous derivations for the one-loop
open and
unoriented string graphs, as well as the supersymmetric extensions,
and in the presence of background two-form fields.
The two-form background field dependence in generic one-loop scattering
amplitudes of the open bosonic string theory was first derived in a paper
with Novak \cite{ncom}. We are only including
calculations pertinent to the macroscopic loop amplitude in this paper.} 

\vskip 0.1in The eigenspectrum of the scalar Laplacian on a
surface with boundary includes a dependence on an electromagnetic
background, reflected as a twist in the boundary conditions
satisfied by the scalar. As an illustration, let us work out the
functional determinant of the scalar Laplacian for worldsheets
with the topology of an annulus with generic twist $\alpha$. We
begin with a discussion of the eigenspectrum. In the case of the
free Neumann scalars, we must introduce an infrared regulator mass
for the zero mode, as in the case of the torus \cite{poltorus}.
The functional determinant of the Laplacian can be written in the
form:
\begin{equation}
{\rm ln} ~ {\rm det}^{\prime} \Delta = \lim_{m \to 0}
  \sum_{n_1 =0}^{\infty} \sum_{n_2 = -\infty }^{\infty}
  {\rm log} \left [ {{\pi^2}\over{t^2}} ( n_2^2 + n_1^2 t^2 + m^2  ) \right ]
 -  {\rm log} \left ( {{\pi^2 m^2}\over{t^2}} \right )
\quad . \label{eq:nfproaa}
\end{equation}
The first term in Eq.\ (\ref{eq:nfproaa}) is a special case of the
infinite sum with generic twist, $\alpha$, and the zeta-regulated
result can be obtained by setting $\alpha$$=$$0$ in the generic
calculation which will be derived below. The second term in Eq.\
(\ref{eq:nfproaa}) yields the result:
\begin{equation}
 + \lim_{m \to 0} \lim_{s \to 0} {{d}\over{ds}}
[{{4\pi^2 m^2  }\over{4t^2 }}]^{-s} =  2~ {\rm log} ~ 2t
 - \lim_{m \to 0}
   - 2 ~ {\rm log} ~ (2\pi m )
 \quad .
\label{eq:zerota}
\end{equation}
This contributes the correct power of $2t$ to the measure of the
path integral for a free Neumann scalar.\footnote{The authors of
\cite{cmnp} were careless on this one point in their analysis.
Their result, which appears in Eq.\ (4.5) of Ref.\ \cite{cmnp},
contains an extraneous factor of $t^{-26/2}$ in the measure. This
would be appropriate for $26$ Neumann scalars, but is clearly
absent in an all-Dirichlet bosonic string amplitude \cite{dbrane,polbook}. See the
comment below  Eq.\ (36).} For the Dirichlet scalar,
we must remember to drop the $n_1$$=$$0$ modes from the double sum
above since the sine eigenfunction vanishes for all values of
$\sigma^1$, not only at the boundary. Thus, the $n_1$ summation
begins from $n_1$$=$$1$. The $n_1$$=$$n_2$$=$$0$ term has been
included in the infinite sum by introducing an infrared regulator
mass, $m$, for the zero mode. We will take the limit $m$$\to$$0$
at the end of the calculation.

\vskip 0.1in We begin by expressing the first term in Eq.\
(\ref{eq:nfproaa}) in the equivalent form:
\begin{equation}
S_{\rm ann} = - \lim_{s,m \to 0} {{d}\over{ds}} \left \{
({{\pi^2}\over{t^2 }})^{-s} \sum_{n_1=0}^{\infty}
\sum_{n_2=-\infty}^{\infty } \left [ n_2^2 + n_1^2 t^2  + m^2
\right ]^{-s} \right \} \quad . \label{eq:nfprost}
\end{equation}
Notice that the infinite sums are manifestly convergent for ${\rm
Re}$ $s$ $>$ $1$. The required $s$ $ \to$ $0$ limit can be
obtained by analytic continuation in the variable $s$. The
analogous step for the second term in Eq.\ (\ref{eq:nfproaa})
yields the relation:
\begin{equation}
 + \lim_{m \to 0} \lim_{s \to 0} {{d}\over{ds}}
[{{\pi^2 m^2  }\over{t^2 }}]^{-s} =  2~ {\rm log} ~ t - \lim_{m
\to 0}  2 ~ {\rm log} ~ (\pi m )
 \quad .
\label{eq:zerot}
\end{equation}
The finite term in this expression contributes the overall factor
of $t^2$ to the result for the one-loop vacuum amplitude.

\vskip 0.1in The infinite summation over $n_2$ is carried out
using a Sommerfeld-Watson transform as in \cite{poltorus}. We
invoke the Residue Theorem in giving the following contour
integral representation of the infinite sum as follows:
\begin{equation}
\sum_{n = -\infty }^{\infty} \left [ n^{2} + x^2 \right ]^{-s} =
\oint_{\sum_n {\cal C}_n} {{dz}\over{2\pi i}} \pi {\rm cot} (\pi
z)
  \left ( z^2 + x^2 \right )^{-s}
\quad , \label{eq:contoureigt}
\end{equation}
where ${\cal C}_n$ is a small circle enclosing the pole at
$z$$=$$n$ in the counterclockwise sense. The contours may be
deformed without encountering any new singularities into the pair
of straight line contours, ${\cal C}_{\pm}$, where the line ${\cal
C}_+$ runs from $\infty $$+$$i \epsilon$ to
$-\infty$$+$$i\epsilon$, connecting smoothly to the line ${\cal
C}_-$, which runs from $-\infty$$-$$i\epsilon$ to
$\infty$$-$$i\epsilon$.

\vskip 0.1in Alternatively, we can choose to close the contours,
${\cal C}_{\pm}$, respectively, in the upper, or lower,
half-planes along the circle of infinite radius. Note that the
integrand has additional isolated poles in the complex plane at
the points $z$$=$$\pm$$ix$. We will make the following
substitution in the integrand:
\begin{eqnarray}
{{{\rm Cot}(\pi z)} \over {i}}=&& {{2e^{ i\pi z}}\over{e^{i\pi
z}-e^{-i \pi z}}}
        - 1 \quad , \quad {\rm Im }~z > 0
\nonumber \\
{{{\rm Cot}(\pi z)} \over {i}}=&& {{-2e^{- i\pi z}}\over{e^{i\pi
z}-e^{-i \pi z}}}
        + 1  \quad , \quad {\rm Im }~z < 0
\label{eq:cotid}
\end{eqnarray}
when the contour is to be closed, respectively, in the upper, or
lower, half-plane. This ensures that the integrand is convergent
at all points within the enclosed region other than the isolated
poles. Thus, we obtain the following alternative contour integral
representation of the infinite sum over $n_2$, setting
$x^2$$=$$n_1^2 t^2 $$+$$ m^2$:
\begin{eqnarray}
\sum_{n_2 = -\infty}^{\infty} \left [ n_2^{2} + x^2 \right ]^{-s}
=&& \oint_{{\cal C}_+} dz
 \left [ {{e^{ i\pi z}}\over{e^{i\pi z}-e^{-i \pi z}}} - \half \right ]
\left ( z^2 + x^2 \right )^{-s}
\nonumber \\
&&\quad + \oint_{{\cal C}_-} dz
 \left [ -{{e^{ - i\pi z}}\over{e^{i\pi z}-e^{-i \pi z}}} + \half \right ]
\left [ z^2 + x^2 \right ]^{-s} \quad . \label{eq:contoureiget}
\end{eqnarray}
Note that the contours are required to avoid the branch cuts which
run, respectively, from $+$$ix$ to $+$$i\infty$, and from $-$$ix$
to $-$$i\infty$.

\vskip 0.1in Now consider the case of the twisted Neumann scalar.
There is no need to introduce an infrared regulator in the
presence of a magnetic field since there are no zero modes in the
eigenvalue spectrum. Thus, the analysis of the infinite eigensum
is similar to that for a Dirichlet scalar, other than the
incorporation of twist. We begin with:
\begin{equation}
S_{\rm ann} = - \lim_{s \to 0} {{d}\over{ds}} [{{\pi^2}\over{t^2
}}]^{-s} \sum_{n_2=-\infty}^{\infty\prime } \sum_{n_1 =
1}^{\infty} \left [ n_2^2 + (n_1+\alpha)^2 t^2 ) \right ]^{-s}
\quad , \label{eq:nfpros}
\end{equation}
and identical statements can be made about its convergence
properties as in the zero external field case. The $n_2$ summation
in $S_{\rm ann}$ is carried out using a contour integral
representation  with $x$$=$$(n_1+\alpha)t$. The $n_1$ summation
can be recognized as the Riemann zeta function with two arguments,
\begin{equation}
 \sum_{n_1 = 0}^{\infty}
(n_1 + \alpha )^{-2s+1} t^{-2s+1} \equiv \zeta(2s-1,   \alpha)
t^{-2s + 1} \quad . \label{eq:zetaa}
\end{equation}
Taking the $s$-derivative followed by the $s$$=$$0$ limit gives,
\begin{eqnarray}
\lim_{s\to 0} {{d}\over{ds}}&&
 \zeta(2s-1,  \alpha)t^{-2s+1} B(\half , s- \half )
\nonumber \\
=&&\lim_{s\to 0} {{d}\over{ds}}
 \zeta(2s-1,  \alpha) t^{-2s+1} {{{\rm sin} (\pi s)}\over{{\sqrt{\pi}}}}
\Gamma (1-s) \Gamma (s - \half ) = - 2 \pi t \zeta (-1,  \alpha)
\quad . \label{eq:finttranseigka}
\end{eqnarray}
Substituting the relation
$\zeta(-n,q)$$=$$-B_{n+2}^{\prime}(q)/(n+1)(n+2)$, and combining
the contributions for $q$$=$$\alpha $, and $1$$-$$\alpha$, gives:
\begin{equation}
{{\pi t }\over{3}}  \left [ B_3^{\prime} (1- \alpha) +
B_3^{\prime} ( \alpha)) \right ] = 2 \pi t \left [ \alpha^2 +
{{1}\over{6}} - \alpha \right ] \quad , \label{eq:tota}
\end{equation}
where we have used $B_n^{\prime}(q)$$=$$nB_{n-1}(q)$, and
$B_2(q)$$=$$q^2$$-$$q$$+$${{1}\over{6}}$.

\vskip 0.1in Next, we tackle the non-constant pieces from the
square brackets in the analog of Eq.\ (\ref{eq:contoureiget}). It
is helpful to take the $s$-derivative and the $s$ $\to$ $0$ limit
prior to performing the contour integral. We begin with the
contour integral in the upper half-plane:
\begin{eqnarray}
I_2 ( x, s) =&& \int_{{\cal C}_+} dz
 {{e^{ i\pi z}}\over{e^{i\pi z}-e^{-i \pi z}}}
  \left ( z^2 + x^2 \right )^{-s}
\nonumber \\
        =&&
- 2 {\rm sin} (\pi s) \int_{x }^{\infty} dy
 {{e^{ -\pi y  }}\over{e^{\pi y } -e^{- \pi y } }}
  \left ( y^2 - x^2 \right )^{-s}
         \quad .
\label{eq:twfsumtrans2eig}
\end{eqnarray}
Taking the derivative with respect to $s$, and setting $s$$=$$0$,
gives:
\begin{eqnarray}
{{d}\over{ds}} I_2 ( x, s)|_{s=0} =&& - 2 \pi \int_{ x }^{\infty}
dy
 {{e^{ -\pi y  }}\over{e^{\pi y } -e^{- \pi y } }}
\nonumber \\
=&& - {\rm log} \left ( 1 +e^{-2\pi x} \right ) \quad .
\label{eq:twequaeig}
\end{eqnarray}
The ${\cal C}_-$ integral gives an identical contribution as
before. Combining the contributions to $S_{\rm ann}$ from terms
with $\alpha$$=$$\phi/\pi$, and $1$$-$$\alpha$, respectively,
gives the following result in the $m$$\to$$0$ limit:
\begin{equation}
S_{\rm ann} =  2 \pi t ( \alpha^2 + {{1 }\over{6}} - \alpha ) - 2
\sum_{n_1=0}^{\infty} {\rm log} \left [ (1+ e^{-2 \pi (n_1 +
\alpha ) t })
 ( 1+ e^{-2 \pi (n_1 + 1 - \alpha ) t }) \right ]
\quad , \label{eq:equsum}
\end{equation}
The result for the functional determinant of the Laplacian acting
on a twisted complex scalar takes the form:
\begin{equation}
\left [ \prod_{\pm} \prod_{n_1=0}^{\infty}
\prod_{n_2=-\infty}^{\infty \prime}
 \omega_{n_2 , n_1 } \right ]^{-1}
= q^{ \half ( \alpha^2 + {{1}\over{6}} -  \alpha) }
 (1-q^{\alpha})^{-1}
 \prod_{n_1=1}^{\infty}
\left [ ( 1- q^{n_1  - \alpha  })( 1- q^{n_1 + \alpha  } ) \right
]^{-1} \quad , \label{eq:result}
\end{equation}
where $q$$=$$e^{-2\pi t}$. The result can be expressed in terms of
the Jacobi theta function as follows:
\begin{equation}
{{q^{\half \alpha^2 +{{1}\over{24}} - {{1}\over{8}}} }\over{ - 2i
{\rm Sin} (\pi t \alpha/2)}}
 \prod_{n_1=1}^{\infty}
\left [ ( 1- q^{n_1  - \alpha  })( 1- q^{n_1 + \alpha  } ) \right
]^{-1}
 = - i {{ e^{\pi t \alpha^2} \eta(it)}\over{ \Theta_{11} (it \alpha ,it) }}
\quad , \label{eq:resulttat}
\end{equation}
with $\alpha$$=$$\phi/\pi$.

\vskip 0.1in Setting $\alpha$$=$$0$ in this expression, and
combining with the result in Eq.\ (\ref{eq:zerota}), gives the
functional determinant of the Laplacian acting on a free Neumann
scalar:
\begin{equation}
\left ( \prod_{n_2=-\infty}^{\infty \prime} \prod_{n_1 =
0}^{\infty} \omega_{n_1 n_2} \right )^{-1/2} = \left
({{1}\over{2t}} \right )
 q^{ -{{1}\over{24}}}
\prod_{n_1=1}^{\infty} ( 1- q^{n_1 } )^{-1} = {{1}\over{2t}} \left
[ \eta (it) \right ]^{-1} \quad , \label{eq:resultd}
\end{equation}
where $q$$=$$e^{-2\pi t}$. The expression for the Dirichlet
determninant is identical except for the absence of the overall
factor of $1/2t$.


\begin{thebibliography}{99}
\bibitem{dbrane}J. Polchinski, {\em Dirichlet-Branes and Ramond-Ramond Charges},
Phys.\ Rev.\ Lett.\ {\bf 75} (1995) 4724.
\bibitem{witdual}E.\ Witten, {\em String Theory Dynamics in Various
Dimensions}, Nucl.\ Phys.\ {\bf B443} (1995) 85, hep-th/9503024.
\bibitem{smallinst}E.\ Witten, {\em Small Instantons in String
Theory}, Nucl.\ Phys.\ {\bf B460} (1996) 541, hep-th/9511030.
\bibitem{witncom}E.\ Witten, {\em Bound States of Strings and
pbranes}, Nucl.\ Phys.\ {\bf B460} (1996) 335, hep-th/9510135.
\bibitem{west}I.\ Campbell and P.\ West, {\em N=2 10d Nonchiral Supergravity
and its spontaneous Compactification}, Nucl. Phys. B243 (1984) 112; 
M. Huq and M. Namazie, 
{\em KaluzaÐKlein supergravity in ten dimensions}, Class. Q. Grav. 2 (1985).; F. Giani 
and M. Pernici, {\em N = 2 supergravity in ten dimensions}, Phys. Rev. D30 (1984) 325.
\bibitem{hulltown}C. M. Hull, P. K. Townsend, {\em Unity of Superstring Dualities},
 Nucl. Phys. B438 
(1995), 109, hep-th/9410167.
\bibitem{pbrane}G.\ Gibbons and K.\ Maeda, {\em Black Holes and Membranes
in Higher Dimensional Theories with Dilaton Fields}, Nucl.\ Phys.\
{\bf B298} (1988) 741. A.\ Dabholkar, G.\ Gibbons, J.\ Harvey, and
F.\ Ruiz-Ruiz, {\em Superstrings and Solitons}, Nucl.\ Phys.\ {\bf
B340} (1990) 33. G.\ Horowitz and A.\ Strominger, {\em Black
Strings and Branes}, Nucl.\ Phys.\ {\bf B360} (1991) 197. M\ Duff
and H.\ Lu, {\em The Self-dual type IIB superthreebrane}, Phys.\
Lett.\ {\bf B273} (1991) 409; {\em Elementary Fivebrane Solutions
of D=10 Supergravity}, Nucl.\ Phys.\ {\bf B354} (1991) 141; {\em
Black and Super pbranes in Various Dimensions}, Nucl.\ Phys.\ {\bf
B416} (1994) 301. P.\ Townsend, {\em p-brane Democracy},
hep-th/9507048.
\bibitem{teit}R.\
Savit, {\em Duality in Field Theory and Statistical Systems},
Rev.\ Mod.\ Phys.\ {\bf 52} (1980) 452, for early discussions and
citations on flux quantization and duality for extended objects in
lattice field theories. P.\ Orland, {\em Instantons and Disorder
in Antisymmetric Tensor Gauge Fields}, Nucl.\ Phys.\ {\bf
B205}[FS5] (1982) 107. R.\ Nepomechie, {\em Magnetic Monopoles
from Antisymmetric Tensor Gauge Fields}, Phys.\ Rev.\ {\bf D31}
(1985) 1921. C.\ Teitelboim, {\em Monopoles of Higher Rank},
Phys.\ Lett.\ {\bf B167} (1986) 63.
\bibitem{romans}L. Romans, {\em Massive N=2A Supergravity in Ten Dimensions},
Phys.\ Lett.\ {\bf B169} (1986) 374.
\bibitem{ps}J.\ Polchinski and A.\ Strominger, {\em New Vacua for Type II
String Theory}, Phys.\ Lett.\ {\bf B388} (1996) 736.
\bibitem{bergt}E. Bergshoeff, M. De Roo, M. Green, G. Papadopoulos, P. Townsend,
{\em Duality of type II 7-branes and 8-branes}, Nucl.\ Phys.\ {\bf
B470} 113.
\bibitem{jmo}B. Janssen, P. Meessen, and T. Ortin,
{\em The D8brane Tied Up: String and Brane Solutions in Massive
Type IIA Supergravity}, Phys.\ Lett.\ {\bf B453} (1999) 229.
\bibitem{troost}M. Massar and J. Troost, {\em D0-D8-F1 in Massive IIA Sugra}, 
Phys.\ Lett.\ {\bf B458} (1999) 283.
\bibitem{polwit}J.\ Polchinski and E.\ Witten, {\em Evidence for Heterotic-Type I
String Duality}, Nucl.\ Phys.\ {\bf B460} (1996) 525.
\bibitem{polyakov}A.\ M.\ Polyakov, {\em Quantum Geometry of
Bosonic Strings}, Phys.\ Lett.\ {\bf B103} (1982) 107.
\bibitem{hawk}S.\ W.\ Hawking, {\em Quantum Gravity and Path
Integrals}, Phys.\ Rev.\ {\bf D18} (1978) 1747. {\em Zeta Function
Regularization of Path Integrals}, Comm.\ Math.\ Phys.\ {\bf 55}
(1977) 133.
\bibitem{poltorus}J. Polchinski,
{\em Evaluation of the One Loop String Path Integral}, Comm. Math.
Phys. {\bf 104} (1986) 37.
\bibitem{cmnp}A.\ Cohen, G.\ Moore, P.\ Nelson, and J.\ Polchinski,
{\em An Off-shell Propagator for String Theory}, Nucl.\ Phys.\
{\bf B267} (1986) 143.
\bibitem{call}A.\
Abouesaood, C.\ Callan, C.\ Nappi, and S.\ Yost, {\em Open Strings
in Background Gauge Fields}, Nucl.\ Phys.\ {\bf B280} [FS18]
(1987) 599.
\bibitem{bachas}E.\
Fradkin and A.\ Tseytlin, Phys.\ Lett.\ {\bf B163} (1985) 123. C.\
Bachas, {\em Dbrane Dynamics}, Phys.\ Lett.\ {\bf B374} (1996) 37.
M.\ Berkooz, M.\ Douglas, and R.\ Leigh, {\em Branes Intersecting
at Angles}, Nucl.\ Phys.\ {\bf B480} (1996) 265. 
N. Seiberg
and E. Witten, {\em String Theory and Noncommutative Geometry},
hep-th/9908142, and references within.
\bibitem{dkps} M.\ Douglas, D.\ Kabat, P.\
Pouliot, and S.\ Shenker, {\em Dbranes and Short Distances in
String Theory}, Nucl.\ Phys.\ {\bf B485} (1997) 85.
\bibitem{polbook}J.\ Polchinski, {\em String Theory}, in two
volumes (1998), Cambridge.
\bibitem{nal}S.\ Chaudhuri, {\em Path Integral Evaluation of
Dbrane Amplitudes}, Phys.\ Rev.\ {\bf D60} (1999) 106007,
hep-th/9907179.
\bibitem{wils}S.\ Chaudhuri, Y.\ Chen, and E.\ Novak, {\em Pair Correlation
Function of Wilson Loops}, Phys.\ Rev.\ {\bf D62} (2000) 026004.
\bibitem{ferm}
S.\ Chaudhuri and E.\ Novak, {\em Supersymmetric Pair Correlation
Function of Wilson Loops}, Phys.\ Rev.\ {\bf D62} (2000) 046002.
\bibitem{flux}S.\ Chaudhuri, {\em Confinement and the Short Type I$^{\prime}$ Flux Tube},
Nucl.\ Phys.\ {\bf B591} (2000) 243, hep-th/0007056.
\bibitem{ncom}S.\ Chaudhuri and
E.\ Novak, 
{\em Effective String Tension and Renormalizability: String Theory 
in a Noncommutative Space}, JHEP 0008 (2000) 027, hep-th/0006014.
\bibitem{zeta}S.\ Chaudhuri, {\em The Normalization of
Perturbative String Amplitudes: Weyl Covariance and Zeta Function
Regularization}, electronic pedagogical review paper,
hep-th/0409031.
\bibitem{bergc}E.\ Bergshoeff, U.\ Gran, R.\ Linares, M.\ Nielsen,
and D.\ Roest, {\em Domain Walls and the Creation of Strings},
hep-th/0303253.
\bibitem{mtheory}S.\ Chaudhuri,
{\em Spacetime Reduction of Large N Flavor Models: A Fundamental
Theory of Emergent Local Geometry?}, Nucl.\ Phys.\ {\bf B719} (2005) 188, hep-th/0408057.
\bibitem{landscape}S.\ Chaudhuri, {\em A Renormalizable Landscape of
Sixteen Supercharges: Electric-Magnetic Duality, Matrices, and Emergent 
Spacetime}, hep-th/0507116.
\end{thebibliography}
\end{document}